# Apparition de structures tourbillonnaires de type Görtler dans une cavité parallélépipédique ouverte de forme variable

**Thierry M. Faure, Nicolas Pechlivanian, François Lusseyran, Luc Pastur**

*Université Pierre et Marie Curie, Paris 6 et Université Paris Sud 11*
*LIMSI-CNRS, UPR 3251*
*B.P. 133 91403 Orsay Cedex*
*email : thierry.faure@limsi.fr*

## Résumé :

*L'interaction entre une couche limite laminaire et une cavité ouverte est étudiée expérimentalement pour des nombres de Reynolds compris entre 860 et 32300. Des visualisations de l'écoulement sont réalisées dans différents plans d'observation afin de comprendre l'évolution spatiale des structures dynamiques. L'étude est effectuée en changeant la longueur et la hauteur de la cavité ainsi que la vitesse de l'écoulement afin de modifier la nature des structures présentes à l'intérieur de la cavité. L'objectif est d'appréhender l'évolution tridimensionnelle de l'écoulement. On montre que les structures dynamiques de la cavité ne sont pas dues à l'évolution des instabilités secondaires de la couche de cisaillement. Des tourbillons de type Görtler apparaissent dans la cavité pour certains paramètres et on analyse leurs propriétés. Le diagramme d'existence de ces instabilités est également établi.*

## Abstract :

*The interaction between a laminar boundary layer and an open cavity is investigated experimentally for Reynolds numbers between 860 and 32300. Flow visualizations are carried out for different observation directions in order to understand the spatial development of dynamical structures. The study is conducted by changing the cavity length and height and the external flow velocity, and therefore the flow patterns inside the cavity. The issue is to emphasize the three-dimensional development of the flow. In particular, we show that the cavity dynamical structures are not due to secondary shear layer instabilities. Görtler-type vortices are developing inside the cavity for some parameters and their properties are discussed. The existence diagram of these instabilities is also provided.*

## 1 Introduction

Les tourbillons de Görtler sont des instabilités de couche limite induites par la courbure de la paroi qui sont à l'origine d'une transition vers la turbulence (Saric, 1994). Cette étude considère un écoulement d'air de viscosité cinématique $\nu$ et une courbure de l'écoulement induite par une cavité de forme parallélépipédique. Elle présente des résultats issus de visualisations dans des plans à l'intérieur de la cavité pour différentes longueurs et hauteurs et pour des nombres de Reynolds, $Re = U_e L / \nu$, compris entre 860 et 32300. L'objectif est de fournir la morphologie de l'écoulement en fonction de la géométrie de la cavité et de s'attacher en particulier aux caractéristiques de tourbillons de type Görtler présents dans certains cas. Les seuils d'apparition et de disparition de ces structures ainsi que leur répartition seront plus particulièrement analysés.

## 2 Dispositif expérimental

L'écoulement est crée par un ventilateur centrifuge placé en amont de la chambre de tranquillisation (Figure 1-a). L'injection de marqueur s'effectue en entrée du ventilateur. Un conduit se terminant par du nid d'abeille et un convergent amène l'écoulement vers la section d'essais, constituée d'une plaque plane munie d'un bord d'attaque elliptique afin de fixer l'origine de la couche limite. La longueur de la plaque $A = 300$ mm permet de fournir une couche limite laminaire établie. Les réflexions lumineuses dues aux parois sont minimisées par l'utilisation d'un verre antireflet de 2 mm d'épaisseur pour l'ensemble de la section d'essais. La hauteur $H$ de la cavité varie entre 25 mm et 150 mm, sa longueur $L$ entre 18,75 mm et 200 mm et son envergure $S = 300$ mm est constante. Les extrémités de la cavité selon cette direction sont les parois verticales de la soufflerie (Figure 1-b). Le rapport d'envergure $F = S / H$ est compris entre 2 et 12 et le rapport de forme $R = L / H$ entre 0,25 et 2,5. En sortie de la soufflerie, l'air est rejeté dans la salle de mesure. La vitesse extérieure $U_e$ est mesurée par vélocimétrie laser Doppler (LDV) 102 mm en amont de la cavité et 25,5 mm au-dessus de la plaque plane. Ce point de mesure est situé dans l'écoulement extérieur suffisamment en amont pour ne subir aucune perturbation de l'instabilité qui se développe au-dessus de la cavité. L'origine du système de coordonnées est prise au bord amont de la cavité à mi-envergure, l'axe $x$ est dans la direction de l'écoulement, l'axe $y$ normal à la plaque amont et l'axe $z$ selon l'envergure. La paroi supérieure de la section d'essais, située à $D = 75$ mm au-dessus de la cavité, n'a aucune influence sur la couche de cisaillement. L'épaisseur de la couche limite sur cette paroi est inférieure à 10 mm et n'a pas d'influence sur l'écoulement extérieur. Les visualisations de l'écoulement sont réalisées dans un plan lumineux obtenu avec la longueur d'onde verte d'un laser argon-ion (514,5 nm). Le marqueur de l'écoulement est de la fumée de spectacle à faible densité.

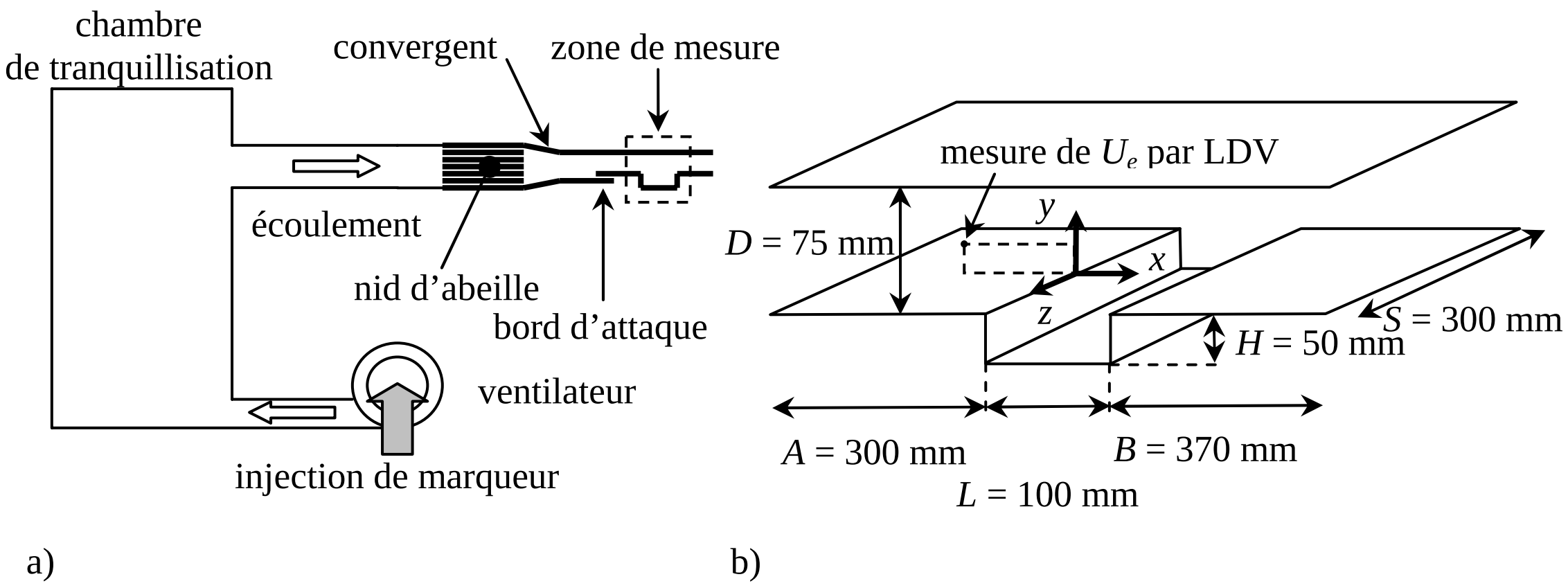

a) b)

Figure 1: Dispositif expérimental : a) soufflerie, b) section d'essais pour un rapport de forme $R = 2$, $F = 6$ et définition du repère.

## 3 Dynamique de l'écoulement

### 3.1 Écoulement dans la cavité

Des visualisations dans un plan vertical ($x$,$y$) mettent en évidence la dynamique de l'écoulement en fonction de la forme de la cavité (Figure 2-a). Un tourbillon s'y développe sous l'effet de l'entrainement engendré par la couche de cisaillement et a pour effet d'induire une courbure de l'écoulement (Faure *et al.* 2005, Faure *et al.* 2006). Ce tourbillon principal peut se limiter à la partie aval de la cavité (par exemple pour $R = 2$, Figure 2-b) la partie amont étant le siège d'un tourbillon secondaire contrarotatif, ou occuper toute la cavité (par exemple pour $R = 1$, Figure 2-c). Alors que la couche de cisaillement présente peu d'évolution selon l'envergure et démontre un caractère bidimensionnel, la dynamique tourbillonnaire dans la cavité est fortement tridimensionnelle (Faure *et al.* 2007).

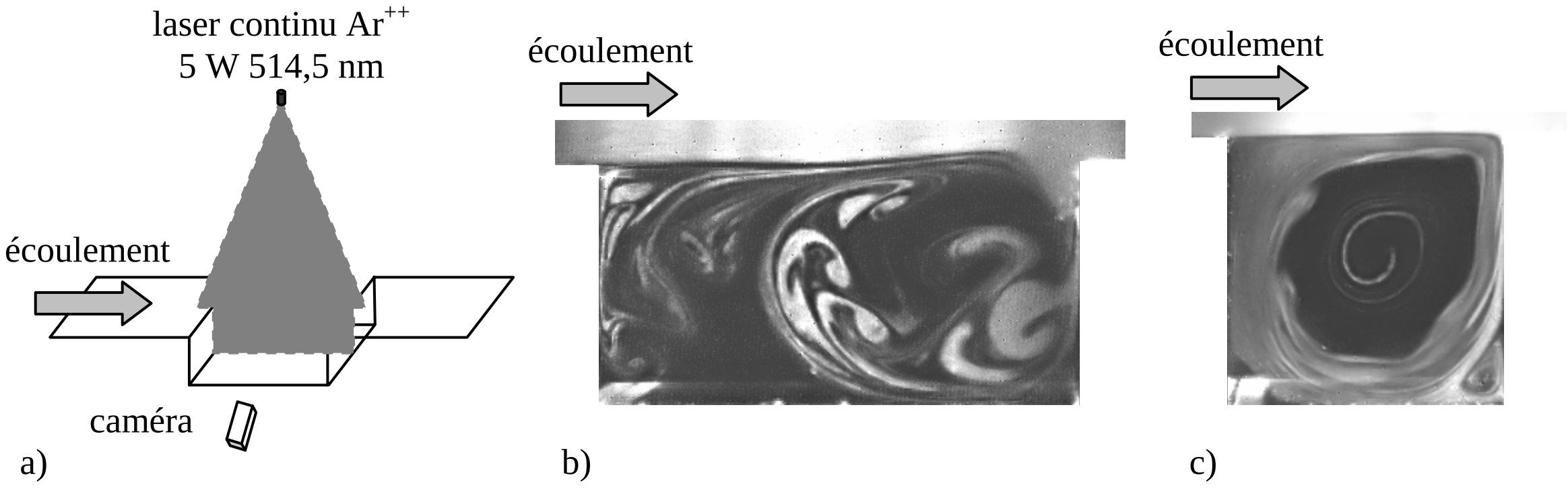

Figure 2: Visualisation dans un plan vertical : a) montage expérimental, b) visualisation pour $R = 2$, $F = 6$ et $Re = 4600$, c) visualisation pour $R = 1$, $F = 6$ et $Re = 4030$.

### 3.2 Développement d'instabilités

La compréhension de la nature tridimensionnelle de l'écoulement est complétée par des visualisations dans un plan horizontal ($x$,$z$) perpendiculaire au plan d'observation précédent (Figure 3-a). La position des plans de visualisation présentés ici est fixée à $y / H = -0{,}3$ par rapport au bord amont de la cavité. On met en évidence un écoulement central, des extrémités latérales vers la ligne médiane de la cavité $z = 0$, qui boucle avec un écoulement de bord, de la ligne médiane vers les extrémités au niveau de la marche amont et de la marche aval de la cavité (Figure 3-b, Faure *et al.* 2007). Les vitesses de cet écoulement transverse sont bien plus faibles que la vitesse de convection du tourbillon principal. Plusieurs morphologies d'écoulement sont observées. L'écoulement transverse décrit précédemment est seul présent pour de faibles nombres de Reynolds (Figure 3-c). Une allée de paires de tourbillons contrarotatifs vient s'ajouter vers le bord amont à l'écoulement précédent à partir d'un nombre de Reynolds seuil (Figure 3-d). Une dynamique de même nature a été observée dans un écoulement de cavité entrainé par un couvercle. Ces instabilités ont été identifiées à des tourbillons de Görtler (Migeon 2002, Guermond *et al.* 2002). Le mécanisme qui leur donne naissance est la courbure de l'écoulement induite par le tourbillon principal. Néanmoins, dans la configuration d'étude présentée ici, où le mécanisme d'entrainement est le cisaillement par l'écoulement extérieur à la cavité, ces instabilités sont de type annulaire comme on peut l'observer sur la Figure 3-d. L'allée située vers le bord amont de la cavité trouve son homologue vers le bord aval. Il faut d'ailleurs ajouter que l'entrainement par l'écoulement transverse est le même pour les structures amont que pour les structures aval, ce qui confirme la forme annulaire des celles-ci. Dans ce cas, il est donc plus approprié de parler d'instabilités de Taylor-Dean pour qualifier ces tourbillons (Mutabazi *et al.* 1989), contrairement à ce qui est observé dans le cas d'un entrainement par une paroi mobile. Un troisième type de comportement est observé pour des nombres de Reynolds supérieurs. Des paires de tourbillons isolées sont présentes par intermittence vers le bord amont ou sont éjectées dans la zone centrale de l'écoulement de la cavité (Figure 3-e). Cette fois-ci, il convient de parler de tourbillons de Görtler au sens strict, puisqu'il n'y a pas de bouclage vers le bord aval (Saric 1994). On observe alors de fortes instationnarités des tourbillons de Görtler suggérant une amorce de transition vers la turbulence.

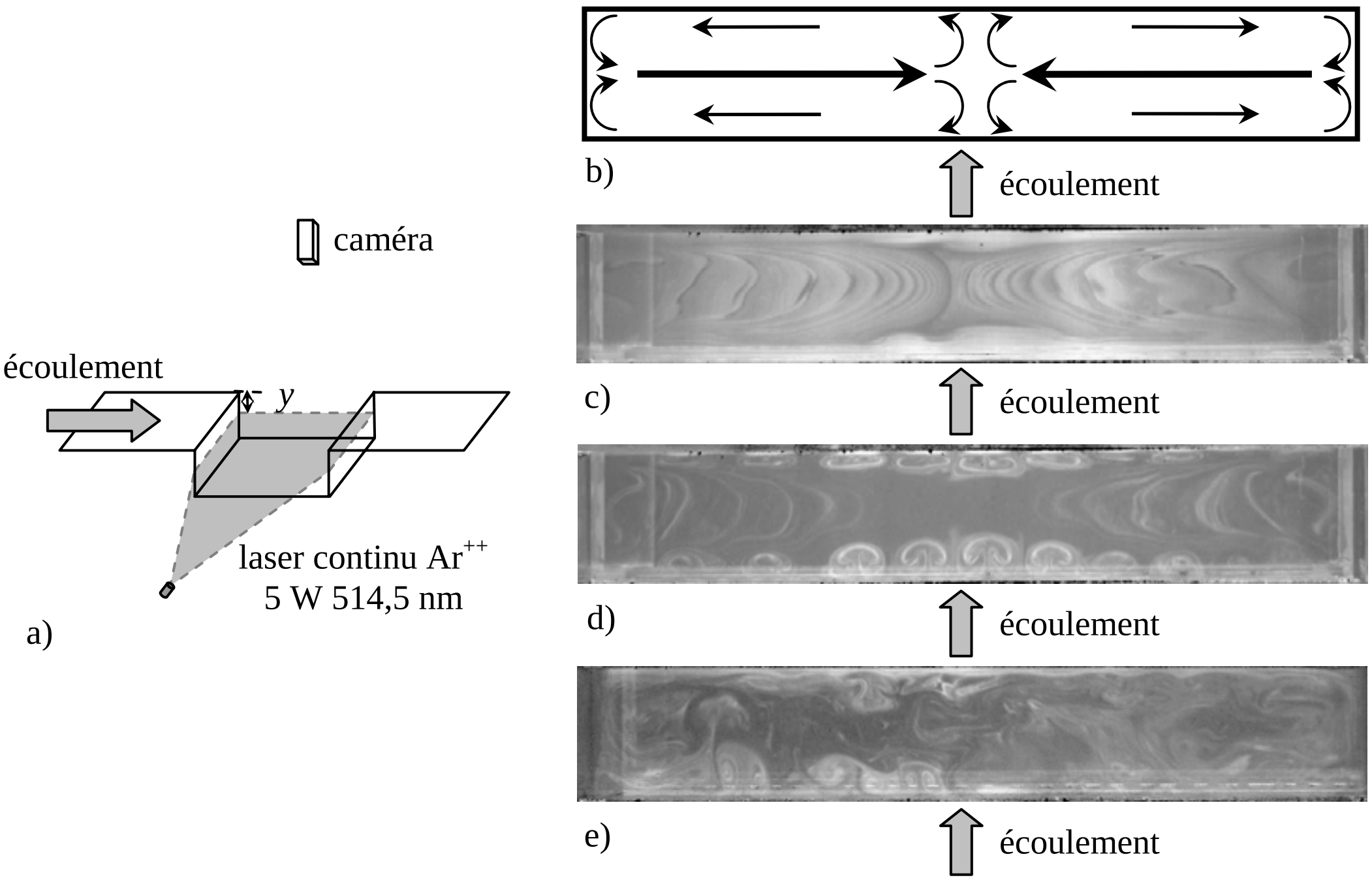

Figure 3: Visualisation dans un plan horizontal : a) montage expérimental, b) schéma de l'écoulement central et des écoulements de bords, c) visualisation pour $R$ = 1, $F$ = 6 et $Re$ = 2970, d) visualisation pour $R$ = 1, $F$ = 6 et $Re$ = 4233, e) visualisation pour $R$ = 0,5, $F$ = 3 et $Re$ = 7500.

### 3.3 Analyse des tourbillons de type Görtler

L'analyse dimensionnelle du problème par le théorème de Vaschy-Buckingham montre que trois nombres sans dimension indépendants sont nécessaires pour décrire complètement l'écoulement de cavité. On retient ici *Re*, *R* et *F* comme groupements adimensionnels. Le domaine d'existence ou d'absence de ces tourbillons en fonction des trois nombres précédents est présenté Figure 4. On remarque ici que la présence de tourbillons considère tout aussi bien le cas des tourbillons de Taylor-Dean se développant en rangée que le cas de tourbillons de Görtler isolés. Pour trois rapports d'envergure, la zone d'existence des tourbillons se présente sous la forme d'un domaine compact avec un seuil d'apparition et un seuil de disparition des tourbillons (Figure 4-a,b,c). Ce n'est pas le cas pour $F$ = 0,5 ou l'on observe deux domaines d'existence des tourbillons dans la gamme d'exploration des paramètres (Figure 4-d). En effet, aucun tourbillon n'est observé le long de la droite $R$ = 0,75 pour $F$ = 2.

L'évolution du nombre de tourbillons est présentée en fonction des nombres adimensionnels Figure 5. Pour chaque rapport d'envergure on observe l'apparition, la présence et la disparition des tourbillons. La géométrie de la cavité est importante puisqu'il existe un rapport de forme pour lequel le nombre de tourbillons de type Görtler atteint un maximum. On remarque Figure 5-a et Figure 5-b que ce nombre maximum de paires de tourbillons est atteint pour $R$ = 1. Cela peut être interprété par la présence d'un tourbillon principal circulaire à l'intérieur de la cavité qui produit une courbure optimale par rapport au confinement des parois. Il y a une chute du nombre de tourbillons pour $F$ = 3 (Figure 5-c) le maximum de tourbillons étant alors obtenu pour $R$ = 0,75. Il faut noter pour ce rapport d'envergure que les tourbillons sont isolés et ne forment pas de rangée. La même remarque est valable pour $F$ = 2 avec un nombre de tourbillons encore plus faible (Figure 5-d). La chute du nombre de tourbillons dans la cavité en fonction du rapport d'envergure montre que le développement de la circulation transverse de fluide est un phénomène qui empêche la formation des tourbillons de Görtler.

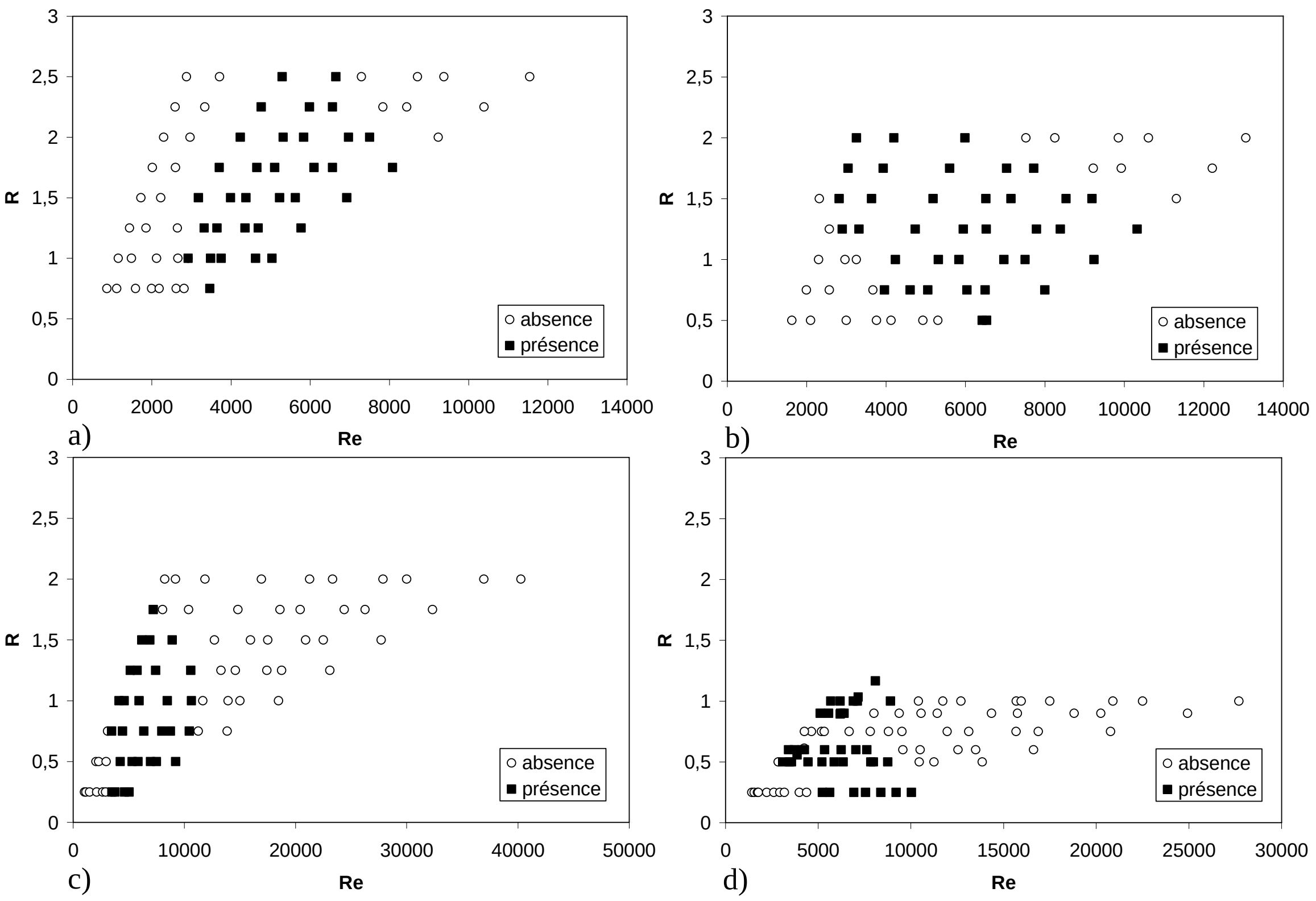

Figure 4: Diagramme de présence ou d'absence des tourbillons de type Görtler en fonction du nombre de Reynolds et du rapport de forme *R* pour : a) *F* = 12, b) *F* = 6, c) *F* = 3 et d) *F* = 2.

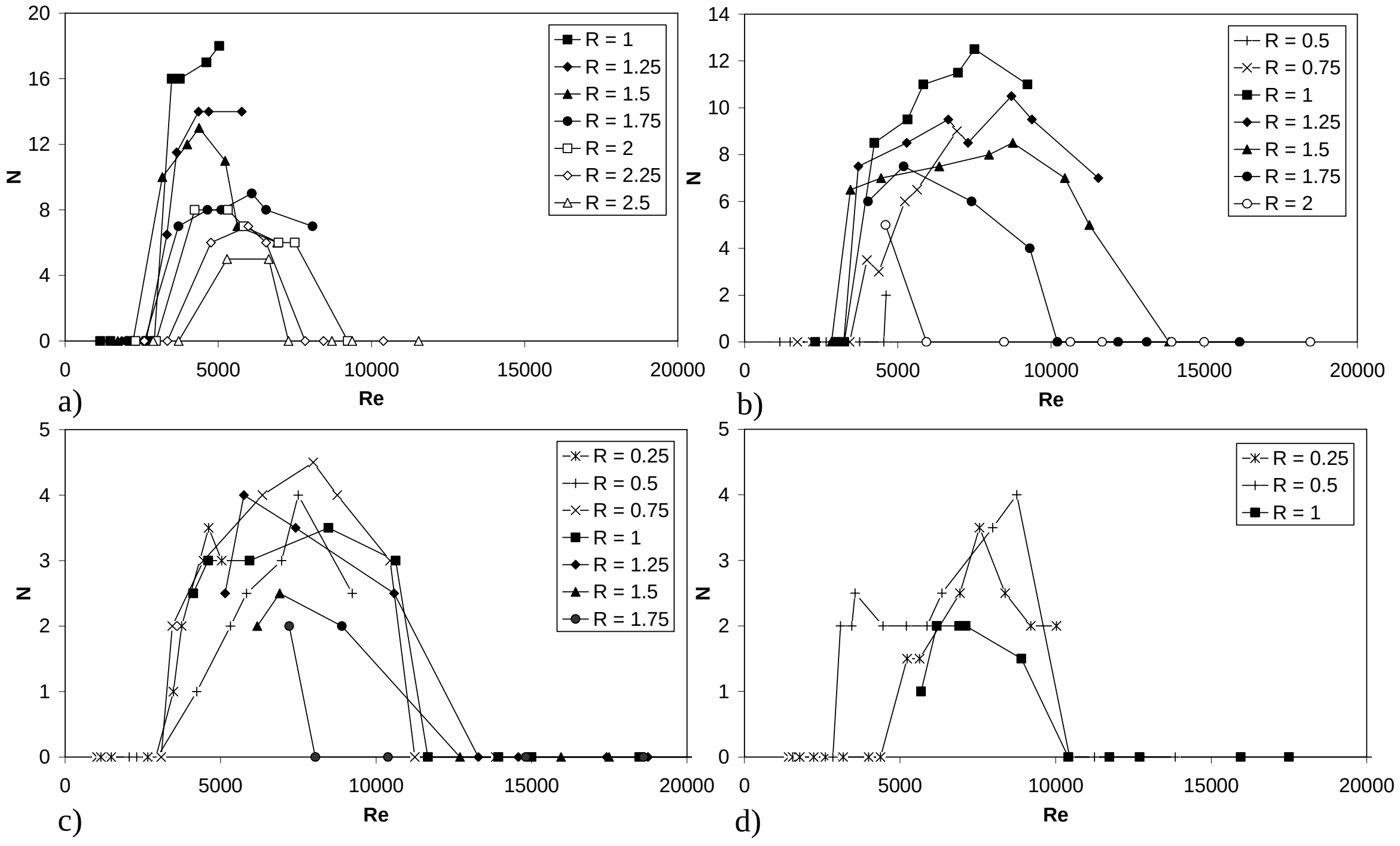

Figure 5: Évolution du nombre *N* de tourbillons de Taylor-Dean à l'intérieur de la cavité en fonction du nombre de Reynolds pour différentes valeurs du rapport de forme *R* et pour : a) *F* = 12, b) *F* = 6, c) *F* = 3 et d) *F* = 2.

## 4 Conclusions

L'écoulement engendré par les oscillations d'une couche de cisaillement au-dessus d'une cavité parallélépipédique met en évidence plusieurs morphologies dominées par un mouvement de convection dû à un tourbillon principal. Des structures tourbillonnaires de type Görtler sont induites par ce mouvement pour certains paramètres de l'écoulement et pour certaines géométries. Ces tourbillons peuvent se développer sous une forme de rangée d'anneaux dans la cavité, correspondant à des tourbillons de Taylor-Dean, ou sous la forme de paires de tourbillons isolés de Görtler au sens propre. Ce comportement diffère du cas d'un écoulement entraîné par un couvercle mobile où il n'y a pas de bouclage des tourbillons sur eux-mêmes. Dans le cas où une rangée de tourbillons est présente dans la cavité et pour un grand rapport d'envergure, un nombre de tourbillons maximum est atteint pour une cavité de section carrée. Ce maximum correspond à un rayon de courbure optimum vis-à-vis du confinement imposé par les parois.

## Références